\documentstyle [12pt]{article}
\def\gammamu{\gamma^{\mu}}
\def\gammanu{\gamma^{\nu}}
\def\g5{\gamma^5}

\def\d4k{{d^4k\over (2\pi)^4}}

\newcommand{\beq}{\begin{eqnarray}}
\newcommand{\eeq}{\end{eqnarray}}
\newcommand{\beqno}{\begin{eqnarray*}}
\newcommand{\eeqno}{\end{eqnarray*}}
\def\lsim{\mathrel{\rlap{\lower4pt\hbox{\hskip1pt$\sim$}}
    \raise1pt\hbox{$<$}}}         
\def\gsim{\mathrel{\rlap{\lower4pt\hbox{\hskip1pt$\sim$}}
    \raise1pt\hbox{$>$}}}         

\begin{document}
\input{psfig}
%
\title{Hadronic Couplings Via QCD Sum Rules Using Three-Point Functions
 : Vacuum Susceptibilities}
 
\author{Mikkel B. Johnson\\
       Los Alamos National Laboratory, \\
       \vspace{5mm}
       Los Alamos, New Mexico\\
        Leonard S. Kisslinger\\
        Department of Physics,\\
       Carnegie Mellon University, Pittsburgh, PA 15213}
\maketitle
\indent
\begin{abstract}
We develop a three-point formalism to treat vacuum susceptibilities used for
the coupling of currents to hadrons within the method of QCD Sum Rules. By 
introducing nonlocal condensates, with the space-time structure taken from 
fits to experimental parton distrbutions, we show that one can treat 
hadronic coupling at zero or low momentum transfer as well as medium and 
asmptotic momentum transfers and obtain a general expression for the vacuum 
susceptibilities of the two-point formalism.  The pion susceptibility, for 
which there has been a major uncertainty, is evaluated successfully with no
new parameters.  
 
\end{abstract}
.
\newpage
\section{Introduction}

Hadronic couplings are essential ingredients in the study of hadronic 
decays and interactions, and the properties and interactions of hadrons in
nuclear matter.  In effective field theories these couplings are defined by 
three-point functions.  Since the hadrons are complex systems and the strong
interactions, given by QCD (Quantum Chromodynamics), require a nonperturbative
treatment, the theoretical treatment of these three-point functions is 
quite challenging.  In the present paper we discuss the application of the
QCD Sum Rule method using a three-point approach for the coupling
of currents to hadrons, and give a new interpretation of the vacuum
susceptibilities used in the two-point approach.  We apply this treatment
to the parity-violating pion-nucleon coupling, for which theoretical estimates
of the pion-induced susceptibility have met with difficulties, and discuss the
isospin-violating pion-nucleon coupling. 

In the method of QCD Sum Rules\cite{svz} complex hadronic systems are
represented by local complex field operators so that standard two-point 
functions can be used for hadronic masses.  The methods introduced by
Shifman et. al. allow a short-distance expansion and nonperturbative effects 
to be treated via operator product expansions (O.P.E.) using
vacuum condensates whose values are determined by fits to experiment, as
well as lattice gauge calculations.
A review of the early work is given in Ref \cite{rry1}.

Using these local field operators, one can also define three-point functions 
for hadronic coupling, similar to effective hadronic field theories.
For medium and asymptotic momentum transfers the O.P.E. can be applied for
form factors\cite{is1,nr} and moments of wave functions [see Ref. \cite{cz}
for review of the early work]; however, at low momentum transfer the O.P.E.
cannot be consistently applied,
as was pointed out in the early work on photon couplings at low momentum
for the nucleon magnetic moments\cite{is,by}.  In Ref.\cite{is} 
the probem was solved by using 
a two-point correlator in an external electromagnetic field, with vacuum
susceptibilities introduced as parameters for nonperturbative propagation
in the external field.  In Ref.\cite{by} a three point formalism was used
with the long-distance effects treated by bilocal corrections; and by
assuming $\rho$-meson dominance results similar to Ref.\cite{is} were
obtained. 
Subsequently the magnetic susceptibility was calculated using the two-point
formalism with extended vector meson dominance model treatments\cite{bk1,
bky} with results similar to the phenomenological treatment of Ref.\cite{is}.
These methods were applied to the study of parton distribution
functions\cite{k,bb} and radiative baryon decay\cite{bbk}, with explicit
treatments of the bilocal operators. A detailed review of the relationship
between the three-point and two-point external field treatments is 
given\cite{bk2} for an extension to nonzero momentum transfer.

The external field method has also been used for the calculation of the
axial coupling constant (g$_A$) \cite{bk,cpw,hhk}, the parity-violating 
pion-nucleon coupling constant (g$_W$)\cite{hhk1} and the nucleon's tensor
charge\cite{hj}  This two-point method,
however, has two main problems: it cannot be used to extend the coupling
to medium and high momentum transfer and there are additional parameters
to be determined: the vacuum susceptibilities.  This latter problem is 
seen to be crucial in the recent
calculation of g$_W$, where a cancellation between perturbative and
nonperturbative contributions is the dominant effect. Moreover, the
phenomenological value obtained for g$_W$
from the study of g$_{\pi N}$, the strong pion-nucleon coupling constant,
differers by as much as an order of magnitude from a theoretical estimate, 
as we discuss in Sec. 2.1.  A three-point method used\cite{rry2}
for an estimate of g$_{\pi N}$ did not use the pion susceptibility.
Also, for the calculation of the nucleon's tensor charge\cite{hj} it has 
been pointed out\cite{bo} that the treatment of the vacuum tensor 
susceptibility is subtle and different treatments can lead to very different
results for the tensor charge.  

Nonlocal condensates have been shown to be useful for representing the 
bilocal vacuum matrix elements needed for the pion wave function\cite{mr} and 
pion form factor\cite{br} over for low to medium momentum transfer.
In this method one does not carry out an O.P.E. for the power corrections
but introduces new phenomenological parameters needed to characterize the
space-time structure of the nonlocal condensates.  The method is simple,
but powerful.  Although new phenomenological parameters are introdued, they
are interesting in themselves. E.g., in a study of parton distribution
functions\cite{jk} the space-time scale of a nonlocal condensate was
determined by a fit to experiment data.

In the present work we start with the standard three-point vertex functions
for hadronic couplings and use nonlocal condensates to represent the bilocal
operators.  
By comparison of terms appearing in the two-point external field expression 
with those in our hybrid expansion of the three-point function, we obtain a
relationship between the nonperturbative elements in the two methods
[(Sect. 2.2)].  From this relationship, it is then possible to obtain the
main  result of this paper, namely an expression for the induced
susceptibilities of the two-point method in terms of well-defined  four-quark
vacuum matrix  elements, and make a simple estimate of their values, using
the estimate of  the space-time structure of the nonlocal quark condensate 
extracted from experimental data on quark distributions. Since the form
assumed for the nonlocal condensates in Ref\cite{jk} does not have
satisfactory analytic properties, we choose a new form and refit the
parameter needed for the present work.
 
In this study we make use of a factorization
of four-quark operators which cannot be extended to the treatment of
hadronic couplings in nuclear media\cite{dl}. Recently,
we have shown\cite{jk1} that the present knowledge of the in-medium $\Delta$
(1232) can constrain the unknown four-quark in-medium condensates.  
In a future publication\cite{jk2} we demonstrate that the study of hadronic 
in-medium couplings using a QCD Sum Rule method with three-point functions
enables us to extend our program.
 
In Sec. 3 we discuss how this method can be used for the study of the 
pion-nucleon coupling, the parity-violating pion coupling to nucleons
and how the gauge-invariant
method for calculating QED corrections in the QCD Sum Rule method\cite{kl}
can be used for determining the QED isospin violations of coupling constants.
Conclusions and discussion is given in Sec. IV.

\section{Coupling Of Currents To Baryons:Three-Point {\em vs.} Two-Point
Formulation }
\hspace{.5cm}
 
In this section we give a discussion of the three-point vs. two-point 
approach for hadronic couplings and show that by introducing the space-time
structure of the condensates one can successfully use the Sum Rule method
to derive new expressions for the induced susceptibilities of the two-point
method. We also discuss the particular problem of the pion susceptibility,
which is the main application of the present paper.
 
Although hadrons are complicated composite systems, both 
in effective hadronic field theories and in the sum rule methods hadrons are
represented by local field operators.
The coupling of a current J$^\Gamma$(y)~=~$ \bar {q}$(y)$ \Gamma q$(y)
to hadrons $\alpha,\beta$ is studied
in such field theories by the three-point function:
\beq
 V^\Gamma_{\beta\alpha}(p,q) & = & \int d^4x \int d^4y e^{ix\cdot p}
e^{-iy\cdot q} <0|T[\eta_\beta(x) J^\Gamma(y)\bar\eta_\alpha(0)]|0> 
\label{eq-3p}
\eeq
where the quantity $\eta_\alpha(x)$ is a field operator representing 
the hadron $\alpha$. In treatments in which QCD and electroweak interactions 
are explicit, as in the QCD Sum Rule method, the
$\eta$ operators must be composite with quark and gluon field constituents,
so that the problem of coupling of currents to hadrons is intrinsically
much more complex than the three-point functions of Eq.(\ref{eq-3p}) for
effective field theories.  In this
section we review how the couplings are represented by three-point
functions and also by two-point functions in the Sum Rule method; and we show
how the vacuum susceptibilities that appear in the two-point method can be
evaluated in terms of four-quark condensates in the three-point approach.
 
\subsection{QCD Sum Rule Two-Point Method For Coupling At Low Momentum}
\hspace{.5cm}
 
  In this subsection we briefly review the two-point effective field 
approach\cite{is} to hadronic couplings and the definitions of vacuum
susceptibilities. In the present work we
discuss only the coupling to nucleons and use as the composite field operator
to represent the nucleon
\beq
\eta(x) & = &\epsilon^{abc}[u^a(x)^T C\gammamu u^b(x)] \g5\gammamu d^c(x),
\nonumber\\
 <0|\eta(x)|proton> & = & \lambda_p v(x),
\label{eq-eta}
\eeq
where C is the charge conjugation operator, the u(x), d(x) are u,d-quark
fields labelled by color, $\lambda_p$ is a structure parameter and $v$(x)
is a Dirac spinor.

   For coupling of the current $J^\Gamma$ to the proton, if one starts with 
$ V^\Gamma(p,q)$ of Eq.(\ref{eq-3p}), for low q there is no justification for
an O.P.E. in the y variable.  This was dicussed at length in the early
three-point function treatment of the nucleons magnetic dipole
moment\cite{by}, but ignored in the treatment\cite{rry2} of the pion coupling
to nucleons and the N-$\Delta$ pionic coupling.  To avoid this difficulty
a two-point formulation of the QCD Sum Rule in an external electromagnetic
field was introduced\cite{is}.  For an external current J$^\Gamma$ the
correlator
\beq
\Pi^\Gamma(p) & = 
  & i\int d^4 e^{ix\cdot p}<0|T[\eta(x)\bar\eta(0)]|0>_{J^\Gamma}
\label{eq-gpi}
\eeq
is used. As can be seen from Eq. (\ref{eq-gpi}) the microscopic evaluation 
of $\Pi^\Gamma(p)$ can be done using the operator product expansion, since
the variable x is at short distance from the origin.  This is done by an
O.P.E. of the quark propagator in the presence of the the $J^\Gamma$ current
\beq
 S^\Gamma_q(x) & = & <0|T[q(x)\bar{q}(0)]|0>_{J^{\Gamma}}, 
\nonumber\\
        & = & S_q^{\Gamma,PT}(x) + S_q^{\Gamma,NP}(x),
\label{eq-gprop}
\eeq
where $S_q^{\Gamma,PT}(x)$ is the quark propagator coupled perturbatively to
the current and $S_q^{\Gamma,NP}(x)$ is the nonperturbative quark propagator
in the presence of the external current, $J^\Gamma$.  
The quantity $S_q^{\Gamma,NP}$(x) can be thought of as
a nonlocal susceptibility; and it is essential to determine the space-time
structure of this susceptibility to predict the coupling at higher momentum
transfer, as we discuss below.  For the two-point treatment at low momentum
transfer the O.P.E. for
$S_q^{\Gamma,NP}$(x) is justified as in the ordinary two-point function,
giving
\beq
  S_q^{\Gamma,NP}(x)  & = & \frac{-\Gamma}{12}<0|:\bar q 
\Gamma q:|0>_{J^{\Gamma}} + \frac{x^2\Gamma}{3 \cdot 2^6} 
<0|:\bar q \sigma\cdot G \Gamma q:|0>_{J^{\Gamma}} +... .
\label{eq-sgnp}
\eeq

Although the O.P.E. can be justified and the sum rules can easily be
derived in this external field two-point method, there is a major problem:
new parameters appear whose determination must be carried out.  For the new
terms in the nonperturbative quark propagator in the external J$^\Gamma$
current, given in Eq.(\ref{eq-sgnp}) and illustrated in Figs. 1e and 1f,
one can write

\hspace{.5cm}
\begin{center}
\centerline{\psfig{figure=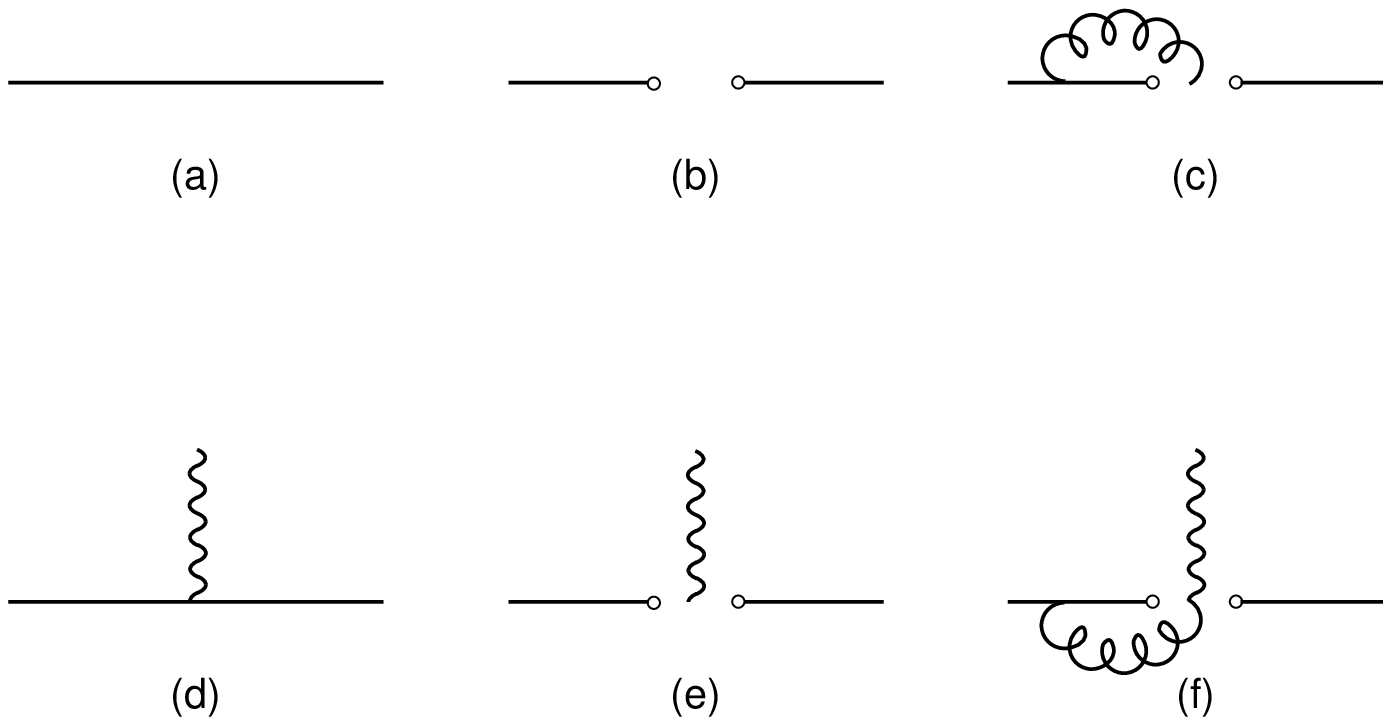,height=6.0cm}}
\vspace{0.1in}
\begin{minipage}{12.0cm} {\small{{\bf Figure 1.} {\em
Diagrammatic representation of terms appearing in the operator-product
expansion of the two-point function in free space (a - c) and
in an external field (d - f)}}}
\end{minipage}
\vspace{0.2cm}
\end{center}

\beq
<0|:\bar{q} \Gamma q:|0>_{J^{\Gamma}} & = & -\chi^\Gamma <0|:\bar{q}q:|0>
\label{eq-chi1}
\eeq
and
\beq
 <0|:\bar{q} \sigma \cdot G  \Gamma q:|0>_{J^{\Gamma}} & = & 
 -\chi^{\Gamma}_{m}<0|:\bar{q}q:|0>
\label{eq-chi2}
\eeq
The
lowest-dimensional diagrams for the microscopic evaluation of $\Pi^\Gamma(p)$
are shown in Fig. 2.  Note that diagrams of Fig. 2b and 2c involve the
susceptibilities $\chi^\Gamma$ and $\chi^\Gamma_m$, respectively.  These
susceptibilities must be determined in order to predict the coupling constant
from the sum rules.  

\hspace{.5cm}
\begin{center}
\centerline{\psfig{figure=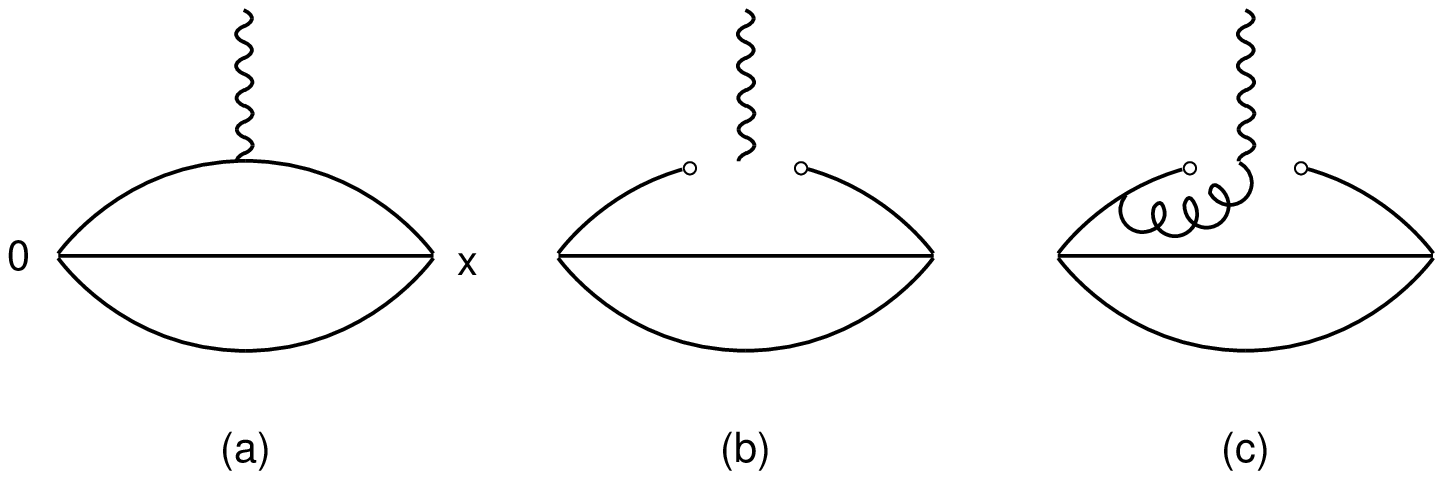,height=4.0cm}}
\vspace{0.1in}
\begin{minipage}{12.0cm} {\small{{\bf Figure 2.} {\em
Lowest dimension diagrams for evaluation of the two-point function
in an external field as given in Eq. (3).}}}
\end{minipage}
\vspace{0.2cm}
\end{center}
 
As an example of the difficulty let us consider the external pion field with
the current J$^\pi$~=~ig$_\pi \bar {q} \tau_3 \gamma_5$q ($\Gamma^\pi$=ig$_\pi
\tau_3 \gamma_5$).  We define the
local pion susceptibility $\chi^\pi$
\beq
<0|:\bar{q} \Gamma^\pi q:|0>_\pi & = & -\chi^\pi  <0|:\bar{q}q:|0>,
\label{eq-chipi}
\eeq
and nonlocal pion susceptibility
\beq
<0|:\bar{q}(x) \Gamma^\pi q(0):|0>_\pi  & = & -\chi^\pi {\rm H(x)}
 <0|:\bar{q}q:|0>.
\label{eq-chinl}
\eeq
The phenomenological function H(x) in Eq.(\ref{eq-chinl}) represents the 
entire O.P.E. of Eq.(\ref{eq-sgnp}). Note that H(0) = 1.
 
A value for the pion susceptibility has been recently extracted\cite{hhk1}
in a study of strong and parity-violating $\pi$-N coupling constant,
g$_{\pi NN}$. The
following problem with the application of PCAC to this problem was observed
in Ref.\cite{hhk1}: The application of PCAC to the determination of the
vacuum pion susceptibility with the two-point method and an external pion
field gives\cite{bk}
\beq
 \chi^\pi <0|:\bar{q}q:|0> & = & \frac{f_\pi^2 m_\pi^2}{\sqrt{2} 2 m_q^2},
\label{eq-pcac1}
\eeq
while from PCAC it is known that
\beq
<0|:\bar{q}(x) \Gamma^\pi q(0):|\pi(k)> & = & -\frac{f_\pi m_\pi^2}{\sqrt{2}
m_q}e^{-ik\cdot x}. 
\label{eq-pcac2}
\eeq
>From Eqs.(\ref{eq-chipi},\ref{eq-pcac1},\ref{eq-pcac2}) it is seen that there
is more than an order of magnitude discrepancy between the two-point
external-field method and standard PCAC, since f$_\pi$/m$_q \simeq$ 20.
In fact the application of Eq.(\ref{eq-pcac1}) gives $\chi_\pi$a $\simeq$
45 GeV$^2$, while the result of the analysis of g$_{\pi NN}$
\beq
 \chi^\pi a & = & 1.88 GeV^2,
\label{eq-chihhk}
\eeq
with a $\equiv$ -(2$\pi$)$^2<0|:\bar{q}q:|0>$. The error in the value of
$\chi^\pi$a is estimated to be about  20\%.  The value of 45 GeV$^2$ is
inconsistent with the sum rules for both the strong and parity-violating
coupling constants, while the value $\chi^\pi$a = 1.88 GeV$^2$ is consistent
with experiment for both the strong and weak coupling.  We derive this
susceptibility in the next section using our three-point method.
 
\subsection{QCD Sum Rule Three-Point Method For Coupling At Low Momentum}
\hspace{.5cm}
 
Let us now return to the three-point function formulation, Eq.(\ref{eq-3p}),
which we write as
\beq
  V^\Gamma(p,q) & = & \int d^4x \int d^4y e^{ix\cdot p} e^{-iy\cdot q} 
 V^\Gamma(x,y) \nonumber\\
 V^\Gamma(x,y) & = & <0|T[\eta(x) J^\Gamma(y)\bar \eta(0)]|0> 
\label{eq-vg2}
\eeq
We write $V^\Gamma(x,y)$ as
\beq
 V^\Gamma(x,y) & = & V^{\Gamma 2q}(x,y) + V^{\Gamma 4q}(x,y)
 +V^{\Gamma 6q}(x,y) + V^{\Gamma 8q}(x,y),
\label{eq-vg3}
\eeq
where the four terms contain two-quark matrix elements only, four-quark,
six-quark and eight-quark matrix elements, respectively. 
Using the
current given by Eq. (\ref{eq-eta}), for which we take 
$\Gamma=g_\pi\gamma_5$ for the pion current, we find for the two-quark terms
\beq
 V^{\Gamma 2q}(x,y) & = & -i2\epsilon^{abc}\epsilon^{b'a'c'}
 \g5 \gamma_\mu S^{ce}_d(x-y) \Gamma S^{ec'}_d(y) \gamma_\nu \g5 
 \nonumber \\
 & & Tr[S^{aa'}_u(x) \gammamu C (S^{bb'}_u(x))^T C\gammanu],
\label{eq-vg2q}
\eeq
which corresponds to Fig. 3a.  The four-quark terms are
\beq
 V^{\Gamma 4q}(x,y) & = & -i 2 \epsilon^{abc}\epsilon^{b'a'c'} 
 <0| \g5 \gamma_\mu d^c(x)\bar d^e(y) \Gamma d^e(y)d^{c'}(0)\gamma_\nu\g5 |0>
  \nonumber \\
 &  & Tr[S^{aa'}_u(x) \gammamu C (S^{bb'}_u(x))^T C\gammanu],
\label{eq-vg4q}
\eeq
where we only show the four-quark condensate term shown in Fig. 3b, since it
is the only term used in the present paper.  We do not consider the six- or
eight-quark condensates in the present work.
 
Note that Fig. 3b for the three-point formulation corresponds to Figs. 2b and
2c  plus the other terms in the O.P.E. for $S_d^{\Gamma,NP}(x)$ of the 
two-point method.  More generally $S_q^{\Gamma,NP}(x)$
for the two-point method is given in the three-point method by
\beq
 S_q^{cc'\Gamma,NP}(x) & = & -i\int d^4y 
 <0|:q^c(x)\bar q^e(y) \Gamma q^e(y)\bar q^{c'}(0):|0>
\label{eq-3p2p}
\eeq
in a linear external field approximation,
where the q$_\mu$ = 0 limit has been taken.  Note that in principle the
space-time structure, as well as the magnitude of the nonlocal susceptibility,
can be determined from the expression Eq.(\ref{eq-3p2p}), and the q$^2$
dependence can be obtained by carrying out the Fourier transform in the 
y-variable.
If we assume vacuum saturation for intermediate states\cite{svz} only the
scalar condensates contribute, and we obtain

\hspace{.5cm}
\begin{center}
\centerline{\psfig{figure=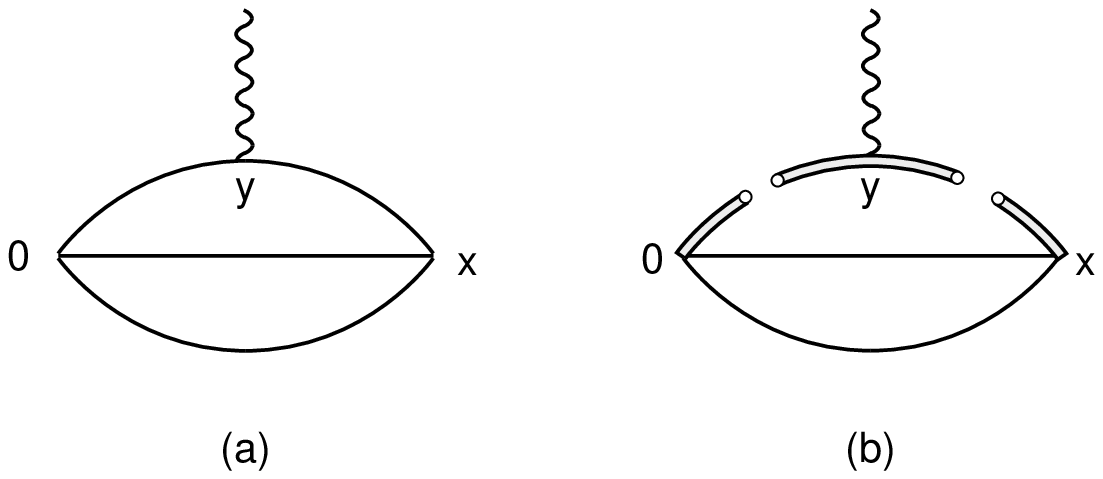,height=4.0cm}}
\vspace{0.1in}
\begin{minipage}{12.0cm} {\small{{\bf Figure 3.} {\em
Two- and four-quark diagrams corresponding to Eqs. (15) and (16), respectively,
for evaluating the coupling constant with the three-point function.}}}
\end{minipage}
\vspace{0.2cm}
\end{center}

\beq
S_{q}^{c c'\Gamma,NP}(x) & \simeq & \Gamma(-i)\int d^4y 
<0|:\bar q^e(y) q^c(x):|0><0|:\bar q^{c'}(0) q^e(y):|0>
\label{eq-4qfac1}
\eeq
In Eq.(\ref{eq-4qfac1}) the nonlocal susceptibility is approximately given 
by nonlocal condensates:
\beq
 <0|:\bar q(0) q(y):|0> & \equiv & {\rm g}(y^2) <0|:\bar q(0) q(0):|0>,
\label{eq-nlc}
\eeq
which gives
\beq
S_{q}^{c c'\Gamma,NP}(x) & \simeq & \Gamma G(x) 
 (<0|:\bar q(0) q(0):|>/12)^2,
\nonumber \\
 G(x) & = & (-i)\int d^4y {\rm g}(y^2) {\rm g}((x-y)^2).
\label{eq-4qfac2}
\eeq
The function g(y$^2$) must be chosen to give satisfactory analytic 
properties as well as consistency with experimental constraints.
Recently, this unknown phenomenological function g(y$^2$) has been fit
to the experimental sea-quark distribution\cite{jk} using a three-point
formulation of deep inelastic scattering in the scaling region. 
For the space-time structure for g(y$^2$) we use
\beq
 {\rm g}(y^2) & = & \frac{1}{(1+\kappa^2y^2/8)^2} 
\nonumber \\
        & = & \int_{0}^{\infty} d\alpha {\rm f}(\alpha)e^{-y^2\alpha/4},
\nonumber \\
 {\rm f}(\alpha) & = & \frac{4}{\kappa^4} \alpha e^{-2\alpha/\kappa^2}.
\label{eq-f}
\eeq
This dipole form is physically reasonable and avoids the undesirable delta
function in the Borel mass which is given by a gaussian form. 
The Jung-Kisslinger monopole form is not satisfactory for the four-quark
nonlocal condensate, but from the range of best fits found in Ref.\cite{jk} 
we estimate that $\kappa^2 \simeq$ (0.15-0.2) GeV$^2$, corresponding to
the quark condensate nonlocality of about 0.2 fm, obtained by
equating the first moment of f($\alpha$) for the dipole form with that
of the monopole form used in Ref.\cite{jk}.  This range of values for 
$\kappa^2$ is obtained by fits to the low-x sea-quark distributions
comparable to those in Ref.\cite{jk}; and the narrowness of the range is
due to the sensitivity to this parameter.
 
Using the form of Eqs.(\ref{eq-nlc},\ref{eq-f}) in Eq.({\ref{eq-4qfac2}) 
we obtain
\beq
 G(x) & \simeq & -\frac{2^7 \pi^2}{\kappa^4 A(A+4)} [ 1-\frac{2+A}
{\sqrt{A^2 + 4 A}} ln (\frac
{\sqrt{A^2 + 4 A} + A}{\sqrt{A^2 + 4 A} - A})],
\label{eq-G}
\eeq
with A=$\kappa^2 x^2/2^3$.
 
Let us apply this to the determination of $\chi^\pi$.  From 
Eqs.(\ref{eq-chinl},\ref{eq-4qfac2},\ref{eq-G}) we find (taking the x=0
limit) that
\beq
 \chi^\pi a & \simeq & \frac{G(0)a^2}{3 \cdot 2^4 \pi^2} \simeq 
\frac{2 a^2}{9 \kappa^4} 
 \nonumber \\
            & \simeq & (1.7 - 3.0) GeV^2,
\label{eq-chipi2}
\eeq
in agreement with the value $\chi^\pi \simeq$ 1.88 GeV$^2$,
found in Ref.\cite{hhk} and discussed in the previous section.
If we use the value $\chi^\pi a \simeq$ 1.88 GeV$^2$, we find that
$\kappa^2 \simeq $ 0.19 GeV$^2$.  Note that although there is about a
20 \% error in the phenomenological value of $\chi^\pi$, the results
are very sensitive to $\kappa$.
 
Finally we would like to point out that from Eqs.(\ref{eq-4qfac2},\ref
{eq-G}) the space-time structure of the nonlocal vacuum susceptibilities 
is given.  This enables one to derive the current-hadron vertices for
low momentum transfer. The method can be immediately extended to medium 
momentum transfer for applications to form factors,
hadronic interactions and so forth, by carrying out the Fourier transform
in the y-variable instead of taking the q=0 limit.
 
\section{QCD Sum Rule Three-Point Method For Parity and Isospin Violations
 Of Pion-Nucleon Vertices}
\hspace{.5cm}
 
The QCD Sum Rule determination of the weak parity-violating and isospin
violating pion-nucleon couplings is done by calculating Z$_0$ and photon
loop corrections to the diagrams used for the strong coupling, some of
which are shown in Fig. 3.  By using a three-point formulation as 
described in the previous section one can carry out this program without
introducing unknown new vacuum susceptibilities to the extent that the
factorization of four quark vacuum matrix elements is justified.  We
briefly describe this procedure. 
 
\subsection{Parity-violating Pion-Nucleon Coupling}
\hspace{.5cm}
 
At the present time experiments have not detected parity-violations
predicted from the one-pion exchange weak interaction.  The parity-violating 
pion-nucleon coupling constant, $f_{\pi NN}$ might be much smaller
than espected from quark models with the standard electroweak theory.
In the Sum Rule approach
the parity-violating pion-nucleon coupling is determined by starting with
$V^\pi(p,q)$ defined by Eq.(\ref{eq-vg2}) with the current J$^\pi$(y)
used for J$^\Gamma$(y) and all Z$_0$ loops included up to the desired
order.  Taking the limit of massive gauge bosons, so that the weak 
interaction becomes a four-fermion interaction with an effective
Hamiltonian
\beq
 H_w & = & \frac{G_F}{2\sqrt{2}} N^\mu N_\mu
\nonumber \\
 N^\mu & = & \bar {q} \gamma^\mu \tau_3 (1-\frac{4}{3}(1+\tau_3)sin^2
 \theta_W -\gamma_5)q,
\label{eq-hw}
\eeq
In this low-energy limit of the Standard Model it was shown in Ref.\cite{hhk1}
that the only nonvanishing weak contributions are in the two spectator quarks,
which are not interacting with the pion field.  The lowest dimensional
diagrams (again without gluon condensates) are shown in Fig. 3.  In the limit
of q$^\mu$ = 0 we find for the three-point function
\beq
 V^\pi (p,q=0) & = & \frac{G_F sin^2\theta_W g_{\pi q}}{3^2 2^8 \pi^6} 
 (\frac{17}{3} - \gamma) [p^6 ln(-p^2) + \frac{4 G(p^2) a^2}{2^4 3 \pi^2}
 p^4 ln(-p^2)].
\label{eq-gpv}
\eeq
This expression can be readily derived from the results of Ref.\cite{hhk1}
and the results of Sec. II of the present paper.
Since this expression includes the entire operator product expression there
is no need to determine the higher-dimensional susceptibilities, such as
the mixed susceptibility of Eq.(\ref{eq-chi2}), which was a significant
uncertainty in the calculation of Ref.\cite{hhk}.  The main result, that
the parity-violating $\pi$-N coupling constant, f$_{\pi NN}$, is much smaller
than expected from quark models, is still valid, but the experimental value
of the strong constant, g$_{\pi NN}$, is not used.  In other words one can
predict both the strong and weak pion-nucleon coupling.
 
\subsection{Isospin-violating Pion-Nucleon Coupling}
\hspace{.5cm}
 
A new analysis of low-energy pion-nucleon scattering data\cite{gak}
that has shown a large isospin violations in the elastic $\pi$-N amplitudes
which are consistent \cite{p} with isospin violations in $\pi$-N coupling
constants.  The QCD Sum Rule calculation with the three-point method is
done as in the calculation of the parity-violating coupling just discussed
with the replacement of H$_W$ by the electromatic interaction and also
including the effects of the current quark mass differences and the 
isospin splitting of the u- and d- condensates.  With the development of
a gauge-invariant theory for electromagnetic corrections in the Sum
Rule method\cite{kl} it is now possible to carry out this calculation.
The calculation is quite complicated, however for the electromagnetic
corrections, which involve the three-loop diagrams resulting from photon
exchange insertions in the diagrams of Fig. 2. These calculations are
being carried out for the octet mass splittings\cite{ckl}, however, and
will be extended to the calculation of the $\pi$-N isospin violations.

\section{Conclusions}
\hspace{.5cm}
 
The three-point function method is usually avoided 
in QCD Sum Rule treatments of
meson-hadron coupling at low momentum transfer Q due to  the fact that
the O.P.E. is valid only at high Q.  There have been extensive previous
studies of the problem of treating long distance bilocal operators for
electromagnetic coupling. In the present work we have shown, the three-point 
function method can be extended to such low-Q processes by introducing
nonlocal condensates, whose parametrizaion has been shown in Ref.\cite{jk} 
to be phenomenologically related to deep inelastic scattering processes.  
The extension of the three-point method in this fashion provides a convenient
method for extending the evaluation
of hadron coupling constants to high dimension without encountering a
divergent O.P.E. expansion.  
 
We applied the three-point method to solve the outstanding problem of
calculating the vacuum succeptibility for pion-nucleon coupling, encountered
in previous applications of the two-point function to this problem.  
We find a vacuum succeptibility of $-\chi^\pi$a = 1.7 - 3.0 GeV$^2$, close to 
the value found in Ref.\cite{hhk1}.  
 
We conclude that the three-point method with nonlocal condenstes to
epresent long-distance effects is a viable approach for calculating 
low momentum-transfer processes in the QCD sum rule approach, and we have 
suggested applications to parity and
isospin violating couplings.  Another application of the three-point method
of great interest is to hadron couplings in nuclei, which will be considered
in a future paper~\cite{jk2}.
 
This work was supported in part by the National Science Foundation grant
PHY-9319641 and in part by the Department of Energy.

\end{document}